\documentclass[fleqn,usenatbib]{mnras}

\usepackage{mathptmx}

\usepackage[T1]{fontenc}

\DeclareRobustCommand{\VAN}[3]{#2}
\let\VANthebibliography\thebibliography
\def\thebibliography{\DeclareRobustCommand{\VAN}[3]{##3}\VANthebibliography}

\usepackage{graphicx}	
\usepackage{amsmath}	
\usepackage{amssymb}	
\usepackage[usenames, dvipsnames]{color}

\newcommand{\fmerge}{$f_{\rm{BH,merge}}$}
\newcommand{\rb}[1]{{\color{Cerulean}{{}}}}
\newcommand{\rjs}[1]{{\color{Green}{{}}}}
\newcommand{\new}[1]{{\color{Black}{{#1}}}}

\title[High spin SMBHs in merger-free galaxies]{Supermassive black holes in merger-free galaxies have higher spins which are preferentially aligned with their host galaxy}

\author[Beckmann, Smethurst et al.]{R. S. Beckmann$^{1}$,\thanks{E-mail: ricarda.beckmann@ast.cam.ac.uk} R. J. Smethurst,$^{2}$,\thanks{E-mail: rebecca.smethurst@physics.ox.ac.uk}\thanks{First-authorship is shared between Beckmann \& Smethurst} B. D. Simmons$^{3}$, A. Coil$^{4}$, Y. Dubois$^{5}$, I. L. Garland$^{3}$, C. J. Lintott$^{2}$, \newauthor G. Martin$^{6,7}$, S. Peirani$^{5,8}$, C. Pichon$^{5}$
\\
$^{1}$Institute of Astronomy and Kavli Institute for Cosmology, University of Cambridge, Madingley Road, Cambridge, CB3 0HA, UK\\
$^{2}$Oxford Astrophysics, Department of Physics, University of Oxford, Denys Wilkinson Building, Keble Road, Oxford, OX1 3RH, UK\\ 
$^{3}$Physics Department, Lancaster University, Lancaster, LA1 4YB, UK\\
$^{4}$ Center for Astrophysics and Space Sciences, University of California, San Diego, 9500 Gilman Dr., MC 0424, La Jolla, CA 92093-0424 \\
$^{5}$ Institut d'Astrophysique de Paris/CNRS, 98 bis blvd Arago 75014 Paris \\
$^{6}$ Steward Observatory, University of Arizona, 933 N. Cherry Ave, Tucson, AZ 85719, USA \\
$^{7}$ Korea Astronomy and Space Science Institute, 776 Daedeokdae-ro, Yuseong-gu, Daejeon 34055, Korea \\
$^{8}$ Universit\'e C\^ote d'Azur, Observatoire de la C\^ote d'Azur, CNRS, Laboratoire Lagrange, Bd de l'Observatoire, CS 34229, 06304 Nice Cedex 4, France \\
}

\date{Accepted 2023 June 9. Received 2023 April 25; in original form 2022 November 24}

\pubyear{2022}

\begin{document}
\label{firstpage}
\pagerange{\pageref{firstpage}--\pageref{lastpage}}
\maketitle

\begin{abstract}
Here we use the Horizon-AGN simulation to test whether the spins of SMBHs in merger-free galaxies are higher. We select samples using an observationally motivated bulge-to-total mass ratio of $<0.1$, along with two simulation-motivated thresholds selecting galaxies which have not undergone a galaxy merger since $z=2$, and those SMBHs with $<10\%$ of their mass due to SMBH mergers. We find higher spins ($>5\sigma$) in all three samples compared to the rest of the population. In addition, we find that SMBHs with their growth dominated by BH mergers following galaxy mergers, are less likely to be aligned with their galaxy spin than those that have grown through accretion in the absence of galaxy mergers ($3.4\sigma$). We discuss the implications this has for the impact of active galactic nuclei (AGN) feedback, finding that merger-free SMBHs spend on average $91\%$ of their lifetimes since $z=2$ in a radio mode of feedback ($88\%$ for merger-dominated galaxies). Given that previous observational and theoretical works have concluded that merger-free processes dominate SMBH-galaxy co-evolution, our results suggest that this co-evolution could be regulated by radio mode AGN feedback.
\end{abstract}




\section{Introduction}
The tight correlations observed between the mass of a galaxy’s supermassive black hole (SMBH) with velocity dispersion \citep{magorrian98, merritt01, hu08, kormendy11a, mcconnell11, vdb16} and stellar bulge mass \citep{marconi03,haringrix04} have long been interpreted as evidence for the merger-driven co-evolution of SMBHs and galaxies. However, recent simulations have shown that less than 35\% of SMBH growth since $z = 3$ is due to accretion triggered by galaxy mergers \citep[possibly less than $15\%$;][]{martin18, mcalpine20}, which challenges the observational paradigm of merger-driven galaxy-SMBH co-evolution regulated by active galactic nuclei (AGN) feedback. To study this observationally, a sample of host galaxies with merger-free evolutionary histories, selected to lack the bulges that develop following major or minor galaxy mergers (\citealt{walker96, hopkins12, tonini16}, including those rare gas-rich mergers after which a stellar disk reforms, e.g. \citealt{sparre17}) can be constructed. Galaxies with bulge-to-total ratios $< 0.1$ are thought not to have had a merger since at least $z \sim 2$ \citep{martig12,martin18} and can be selected observationally to isolate merger-free systems. \cite{ssl17}  showed that 101 unobscured AGN hosted by such disk-dominated galaxies with assumed merger-free evolutionary histories had substantial SMBH masses, with the majority lying well above (up to 2~dex) the stellar bulge mass-SMBH mass correlation that is observed for galaxies with significant merger histories, and lying on the typical total stellar mass-SMBH mass correlation.

Follow-up studies of a subset of this disk-dominated sample by \cite{smethurst19} and \cite{smethurst21} revealed outflows ionised by the AGN from these merger-free systems. AGN feedback is thought to be a key regulator of co-evolution and considered necessary in cosmological volume simulations employing $\Lambda$CDM, yet the role of AGN feedback in the absence of mergers is currently unknown. Merger-free AGN growth and feedback are severely understudied, in part because the bulk of previous observational studies either explicitly examine systems with merger-dominated growth histories \citep[e.g. ULIRGs;][]{tadhunter18, perna21} or focus on samples where the merger-driven and merger-free accretion histories cannot be disentangled (i.e. a mixed morphology sample across the entirety of SDSS such as \citealt{RW18}). For this reason, the processes powering the SMBHs accretion that leads to the AGN outflows revealed by \citet{smethurst19, smethurst21} is still poorly understood. Given that a flurry of new results are suggesting that SMBHs grow predominantly through galaxy merger-free processes (\citealt{ssl17, martin18, smethurst19, mcalpine20, smethurst21};  and see also a companion paper to this work \citealt{smethurst22}), it is imperative that we understand the galaxy-merger-free mechanisms responsible for the majority of SMBH growth and the subsequent AGN feedback. 

One important property of SMBHs influenced by their co-evolution with their host galaxy is the magnitude of the BH spin, along with its orientation with respect to its host galaxy. The spin magnitude determines the efficiency with which the SMBH converts accreted mass to feedback energy, while the spin direction determines where feedback energy is deposited if the SMBH is driving a jet. We do have clear theoretical predictions for the expected distribution of the spins and geometries of SMBHs in systems evolving in the absence of mergers which could power such outflows. \cite{npk12} discussed the implications of merger versus non-merger-driven accretion onto a black hole (BH). Merger-driven accretion occurs chaotically, with material infalling towards the SMBH at random angular momentum vectors, on average spinning down the BH \citep{Berti08,dotti13}. Conversely, material inflowing to the SMBH in the centre of a non-merger grown system will come from within the galactic disk at a constant angular momentum vector, spinning up the SMBH to maximum (\citealt{npk12,dubois14,bustamante19})\footnote{Interestingly, \cite{npk12} concluded from this theoretical consideration that secularly grown SMBHs should be over-massive in comparison to merger grown BHs, which they pointed out was in contradiction to measurements of SMBH masses in galaxies with pseudo-bulges available at the time \citep[e.g.][]{kormendy11a}. However, the work of \citet*{ssl17} and \cite{martin18} has significantly eased this contradiction.}. The Bardeen-Petterson effect is then thought to realign the spin of the BH according to that of the accretion disk formed from infalling gas (and therefore the galactic disk in a system fed by planar accretion; \citealt{bardeen75}). The timescales of this process have been the subject of much discussion \cite[see e.g.][]{REES78,Papaloizou83,Scheuer96,Fragile07,Sorathia13} but are thought to be much shorter than the lifetimes of outflows or jets \citep{natarajan98}. Theoretical understanding therefore suggests that outflows will be produced perpendicular to galactic disks specifically in isolated, merger-free systems, and produced at random orientations in systems which have undergone a merger.

There is very little agreement in the literature over the accuracy of methods that claim to be able to probe the spin of SMBHs observationally (e.g. using X-ray reflection spectroscopy to probe the spin; see review by \citealt{reynolds14}). In addition, although the orientation of outflows has been studied extensively in the literature with many studies finding no correlation with disk orientation \citep[e.g. see ][]{kinney00, schmitt03, ruschel21}, there has not yet been an observational study on the orientation of outflows in purely secularly fed systems, only those with mixed accretion histories. 

We note that this would be possible with the high resolution provided by the Hubble Space Telescope (HST) ramp filters used to isolate outflowing \textsc{[oiii]} emission ionised by the AGN. However, since such data does not yet exist, and given the debate over whether it is observationally possible to determine the SMBH spin with X-ray reflection spectroscopy, we turn to simulations to test our hypothesis of non-merger driven SMBH growth. Following the ideas of \cite{npk12} and the results of \cite{smethurst19},  we test whether SMBHs in merger-free systems have higher spin magnitudes and are subsequently aligned with their galactic disks using the Horizon-AGN simulation\footnote{https://www.horizon-simulation.org/}. Horizon-AGN is a modern,  large-scale galaxy evolution simulation that evolves the evolution of a large sample of galaxies from cosmic dawn to redshift $z=0$. It has been shown to reproduce a wide range of observable properties of the galaxy and BH population, such as the galaxy mass functions and cosmic star formation history, the BH mass and luminosity functions, and correlations between BHs and their host galaxies such as the BH-stellar mass relation. While Horizon-AGN did not track BH spin evolution on the fly, BH spins were post-processed for all BHs in Horizon-AGN and presented in \citet{dubois14}.

We describe the Horizon-AGN simulation in Section \ref{sec:sim}, the calculation of BH spins in \ref{sec:spin_model} and our galaxy sample selection in Section~\ref{sec:sample}. Our results are shown and discussed in Section \ref{sec:results}, and we summarise our conclusions in Section \ref{sec:conclusions}.

\section{Simulation Data}
\subsection{Horizon-AGN simulation}\label{sec:sim}
Horizon-AGN is a cosmological-volume hydrodynamical simulation, which has been described in detail in \cite{Dubois14HAGN}. Here we only reiterate its most important features.

Horizon-AGN was run using the adaptive mesh refinement code \textsc{ramses} \citep{teyssier02}, using a standard $\Lambda$CDM cosmology with total matter density $\Omega_m=0.272$, dark energy density $\Omega_\Lambda = 0.728$, amplitude of the matter power spectrum $\sigma_8 = 0.81$, baryon density $\Omega_b = 0.045$, Hubble constant $H_0 = 70.4 \rm \ km s^{-1} \ Mpc^{-1}$ and spectral index $\rm n_s = 0.967$ using WMAP-7 cosmology \citep{komatsu11}. The simulation box has a size of $L_{\rm box} 100 \rm \ h^{-1} Mpc$ (comoving) and is refined on a root grid of $1024^3$. From here, cells are further adaptively refined up to a maximum resolution of $\Delta x = 1$ proper kpc (level 17). Cells are (de)refined when the mass in a cell is more (less) than 8 times the initial mass resolution. The simulation has a DM mass resolution of $M_{\rm DM} = \sim 8.27 \times 10^7 \rm \ M_\odot$, and includes prescription for gas cooling including the contribution from metal released by SN feedback, background UV heating, star formation and stellar feedback. Star formation is modelled according to a Schmidt law with a 1 percent efficiency, using a Salpeter initial mass function. Stellar feedback is modelled to include stellar winds, type Ia and type II supernovae \citep{Dubois08,Kimm15}. The minimum stellar resolution is $\sim 2 \times 10^6 \rm \ M_\odot$.

BHs are created at $z>1.5$ in cells that exceed the density threshold for star formation ($n_0 = 0.1 \rm \ H cm^{-1}$) with an initial seed mass of $10^5 \rm \ M_\odot$. To avoid multiple black holes (BHs) forming in the same galaxies, a 50 comoving kpc exclusion zone for new BH formation is enforced around each existing BH. BH accretion and feedback is modelled as in \citet{dubois12}. BHs accrete gas via the Bondi-Hoyle-Lyttleton formalism $\dot{M}_{\rm BH} = 4 \pi \alpha G^2 M_{\rm BH}^2 \bar{\rho} / (\bar{\rm c}_{\rm s}^2 + \bar{u}^2)^{3/2} $ where $M_{\rm BH}$ is the BH mass, $G$ is the gravitational constant, and $\bar{\rho}$, $\bar{c}_s$ and $\bar{u}$ are the average gas density, sound speed and gas velocity, $\alpha$ is a dimensionless boost factor. \new{We set $\alpha= n / n_0$ if if the gas number density $n > n_0$, and $\alpha=1$ otherwise \citep{booth10}}. Accretion onto the BH is limited at the Eddington accretion rate $\dot{M}_{\rm Edd}$. The Eddington ratio is defined to $f_{\rm edd}=\dot{M}_{\rm BH}/\dot{M}_{\rm Edd} \leq 1$, and measures the efficiency of BH accretion. BH spin is not followed throughout the simulation and is instead computed in post-processing (see Sec. \ref{sec:spin_model}).

AGN feedback energy is released  at a rate of $\dot{E}_{\rm AGN} = \epsilon_f \epsilon_r \dot{M}_{\rm BH} c^2$ where $\epsilon_r=0.1$ is the assumed radiative efficiency  and $c$ is the speed of light. When $f_{\rm edd} > 0.01$, the AGN is \emph{quasar} mode, and energy is injected isotropically as thermal energy with coupling efficiency $\epsilon_f=0.15$. If $f_{\rm edd} \leq 0.01$, the AGN is in  \emph{radio} mode, where energy is released in bi-conical outflows, $\epsilon_f = 1$,  and drive powerful jets when $10^{-4}  < f_{\rm edd} < 10^{-2}$. BHs in Horizon-AGN are able to move freely within their host galaxy, rather than being pinned to their center. To account for un-resolved dynamical friction forces, a sub-grid force of magnitude $\mathcal{F}_{\rm DM} = f_{\rm gas} 4 \pi \alpha \rho (G M_{\rm BH}/\bar{c}_s)^2$  is added following \cite{ostriker99}. $f_{\rm gas}= 0-2$ \citep{chapon2013} is a factor that depends on the Mach number. BHs merge when located within 4 kpc of each other, and when their relative velocity is smaller than the escape velocity of the binary. 

\subsection{Computing SMBH spin evolution}
\label{sec:spin_model}

BH spin was not included in the Horizon-AGN run. Instead, it is post-processed from quantities recorded at each coarse timestep throughout the simulation. For our analysis, we use the spins calculated in \citet{dubois14}, using an analytic model that estimates the SMBH spin evolution using gas quantities at the resolution scale of the simulation, as well as information on BH-BH mergers. We briefly reiterate the key features of the model here \new{but refer readers to \citet{dubois14} for details}

All BHs are assumed to form with zero spin (BH spin parameter $\rm |a| = 0$). The evolution of BH spin direction and magnitude is integrated throughout its evolution using the BH accretion rate, instantaneous BH mass and BH \new{angular momentum $\mathbf{J}_{\rm BH}$}, and the angular momentum vector of the accreted gas at each timestep recorded throughout the simulation. \new{In brief, the model works as follows: At each timestep, the angular momentum of the gas $\mathbf{J}_{\rm gas}$ to be accreted is measured from the simulation within the accretion region of the BH. This measured angular momentum is only used to determine the direction of angular momentum accreted by the BH, not the magnitude. If the angular momentum of the gas and BH are misaligned, the gas excerpts a torque on the BH spin through the Bardeen-Petterson effect \citep{bardeen75}. In this case, the inner accretion disc will warp and both BH spin and disc momentum will reorient until $\mathbf{J}_{\rm BH}$ and $\mathbf{J}_{\rm gas}$ are aligned (or anti-aligned) with $\mathbf{J}_{\rm tot}=\mathbf{J}_{\rm BH}  + \mathbf{J}_{\rm gas}$, which determines the new direction of the spin vector of the BH. At this point, BH and accretion disc can be either aligned or anti-aligned, which we determine using \citep{king05}. }

\new{The change in BH spin magnitude due to angular momentum accreted from the disc onto the BH is computed following \citet{bardeen70}. Each accretion event is treated as the formation of a new accretion disc.} We do not model the spin-down of BHs through the Blandford-Znajek mechanism \citep{blandford77}, during which BHs can be spun down as spin energy is extracted by BH-driven jets. During BH-BH mergers, the BH spin of the merger remnant is computed from the properties of the primary and secondary BH following \citet{rezzolla08}.


The galaxy spin axis is measured by computing the angular momentum vector $\vec{a}_{\rm{gal}}$ of all star particles associated with the galaxy. The angle, $\phi_{\rm{gal,BH}}$, between the vectors defining the galaxy spin, $\vec{a}_{\rm{gal}}$, and SMBH spin, $\vec{a}_{\rm{BH}}$, in $\{x$, $y$, $z\}$ Cartesian coordinates within the simulation, was calculated as in the inverse cosine of the dot product of the two vectors divided by the product of their magnitudes:
\begin{equation}\label{eq:angle}
    \cos \phi_{\rm{gal,BH}} = \left(\frac{\vec{a}_{\rm{gal}}.\vec{a}_{\rm{BH}}}{|\vec{a}_{\rm{gal}}||\vec{a}_{\rm{BH}}|}\right).
\end{equation}

\begin{figure*}
    \begin{center}
	\includegraphics[width=\textwidth]{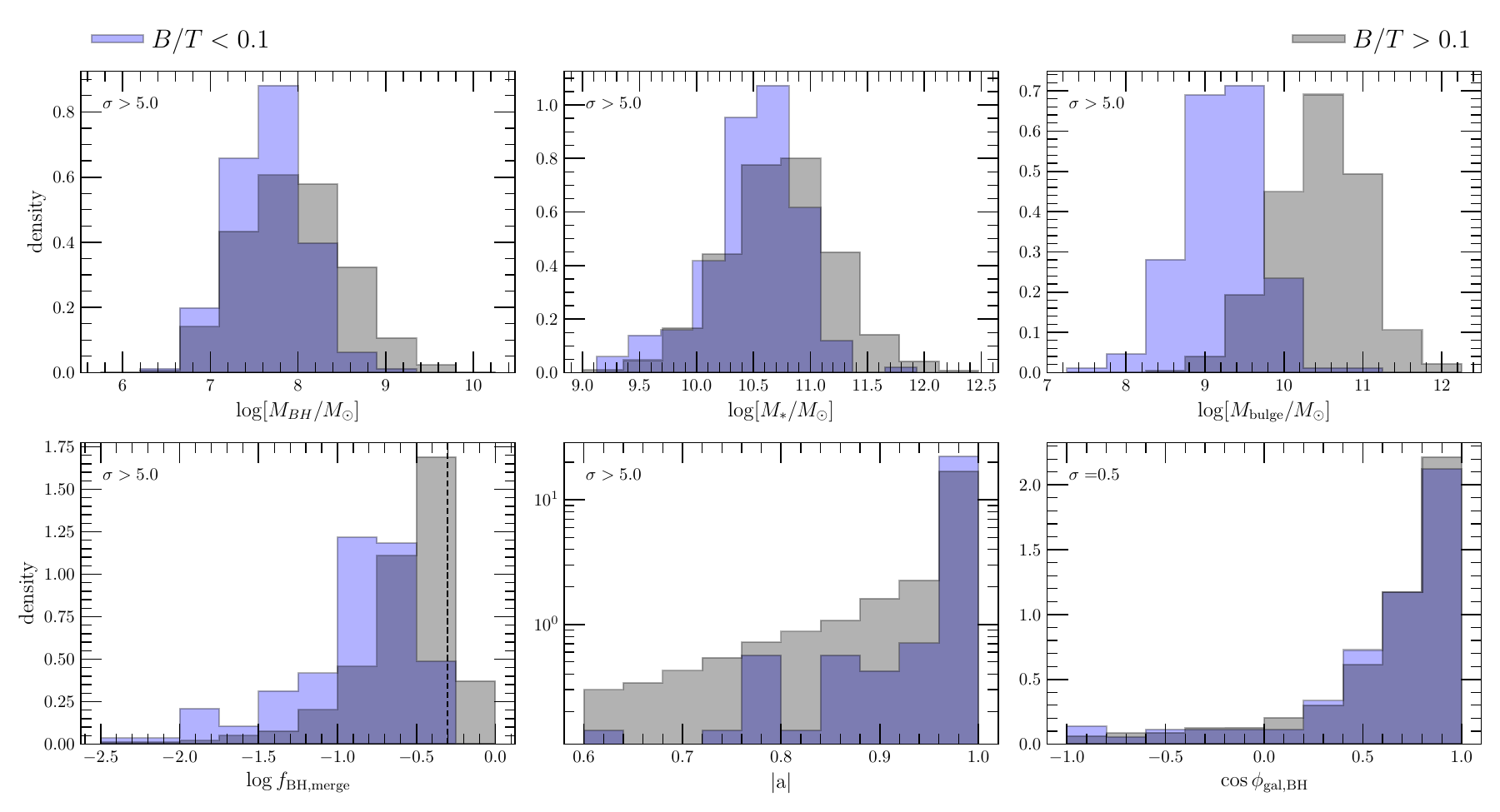}
	\vspace{-1em}
    \caption{The distribution of properties across the Horizon-AGN sample for galaxies selected to have low bulge-to-total ratios, $B/T <0.1$ (blue; assumed to have accretion grown SMBHs), compared to the rest of the sample with $B/T>0.1$ (black; i.e. merger and accretion grown SMBHs). Shown are the SMBH masses, $M_{BH}$, total stellar masses, $M_{*}$, bulge stellar mass, $M_{\rm{bulge}}$, the fraction of the SMBH mass resulting from mergers, \fmerge~ (the dashed line shows \fmerge$ = 0.5$ and denotes the threshold between non-merger on the left and merger dominated growth on the right), the spin magnitude of the SMBH, $|a|$, and the angle between the spin of the SMBH and the galaxy, $\cos\phi_{\rm{gal, BH}}$. A value of $\cos\phi_{\rm{gal, BH}}=1$ means the SMBH and galaxy spins are aligned,  $\cos\phi_{\rm{gal, BH}}=0$ means they are misaligned by $90^{\circ}$, and $\cos\phi_{\rm{gal, BH}}=-1$ means they are misaligned by $180^{\circ}$. A value of $|a|=1$ represents a maximally spinning BH. `Bulgeless' systems with $B/T < 0.1$ have statistically significantly lower SMBH masses, lower stellar masses, lower bulge masses and lower SMBH merger mass fractions. However, they have statistically significantly higher spin magnitudes, yet no statistically significant difference in spin alignment.  We note that these results hold when incomplete, total stellar mass-matched samples of merger-free and merger-dominated systems are compared.}
    \label{fig:BTsplit}
\end{center}
\end{figure*}

\begin{figure*}
    \begin{center}
	\includegraphics[width=\textwidth]{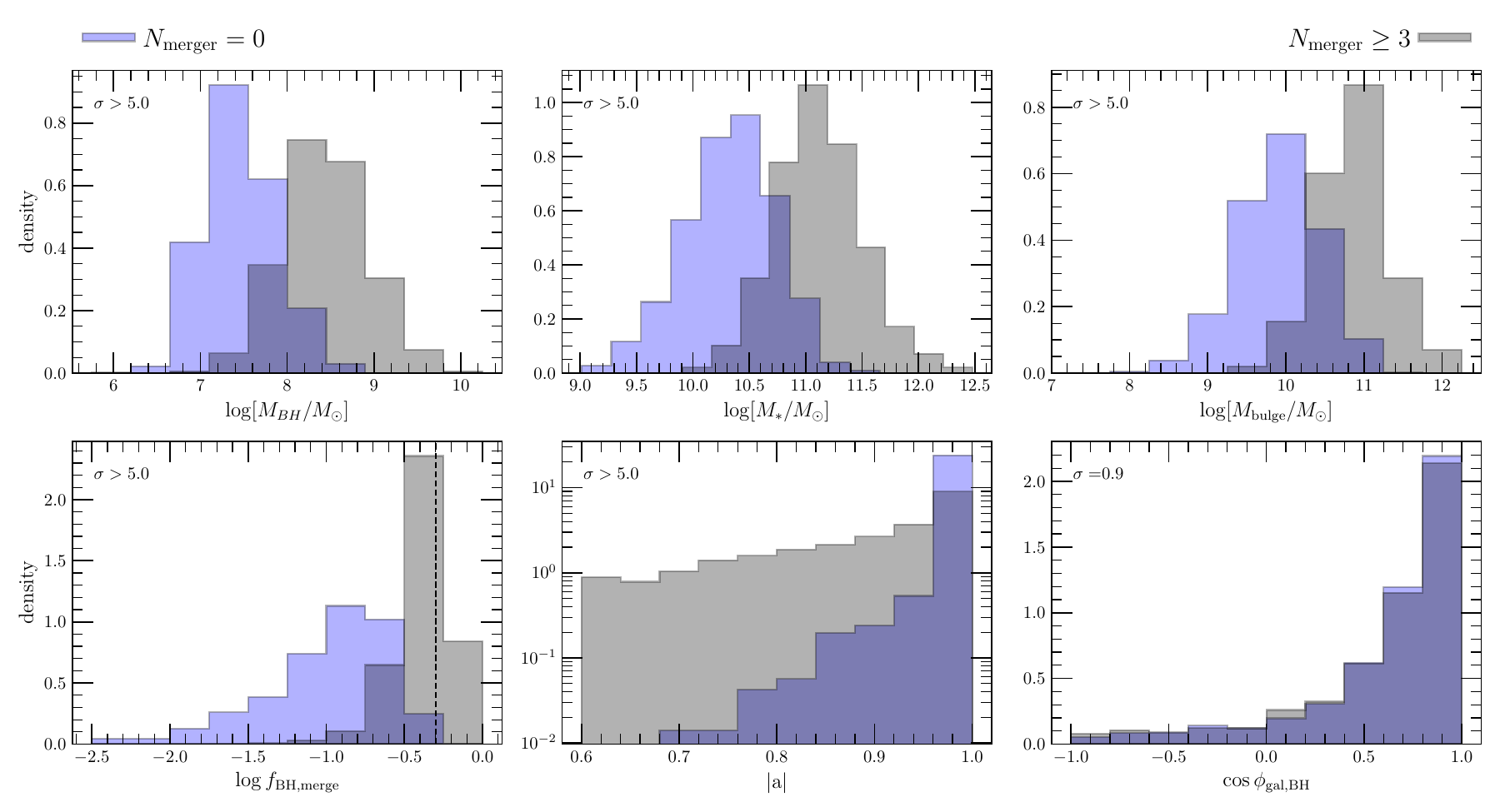}
	\vspace{-1em}
    \caption{The distribution of properties across the Horizon-AGN sample for galaxies selected to have had no major or minor mergers since $z=2$, $N_{\rm{merger}}=0$ (blue; assumed to have accretion grown SMBHs), compared to those which have had more than 3 major or minor mergers since $z=2$, $N_{\rm{merger}}>3$ (black; i.e. merger dominated SMBH growth). Shown are the SMBH masses, $M_{BH}$, total stellar masses, $M_{*}$, bulge stellar mass, $M_{\rm{bulge}}$, the fraction of the SMBH mass resulting from mergers, \fmerge~ (the dashed line shows \fmerge$ = 0.5$ and denotes the threshold between non-merger on the left and merger dominated growth on the right), the spin magnitude of the SMBH, $|a|$, and the angle between the spin of the SMBH and the galaxy, $\cos\phi_{\rm{gal, BH}}$. A value of $\cos\phi_{\rm{gal, BH}}=1$ means the SMBH and galaxy spins are aligned,  $\cos\phi_{\rm{gal, BH}}=0$ means they are misaligned by $90^{\circ}$, and $\cos\phi_{\rm{gal, BH}}=-1$ means they are misaligned by $180^{\circ}$. A value of $|a|=1$ represents a maximally spinning BH. Accretion dominated systems with $N_{\rm{merger}}=0$ have statistically significantly lower SMBH masses, lower stellar masses, lower bulge masses and lower SMBH merger mass fractions. However, they do have statistically significantly higher SMBH spin magnitudes, yet once again no statistically significant difference in spin alignment.  We note that these results hold when incomplete, total stellar mass matched samples of merger-free and merger dominated systems are compared.}
    \label{fig:Nmergesplit}
    \end{center}
\end{figure*}

\subsection{Galaxy sample selection}
\label{sec:sample}

Galaxies and dark matter halos in Horizon-AGN were identified using  \textsc{adaptahop} \citep{aubert04,tweed09}, using a local density threshold of $\rho_{\rm t} = 178 $ times the average dark matter density, the 20 closest neighbours and a force softening of 2 kpc.  A minimum stellar mass cut of $M_{*} = \sim 10^{9} \rm M_\odot$ was applied, where $M_{*}$ is the total mass of all star particles associated with a given galaxy as identified by \textsc{adaptahop}. Bulge-masses are computed as in \citet{volonteri16}, using a first Sersic profile with $n=1$ for the disk component, and a second Sersic profile with the best fit of  $n = 1,2,3$ or $4$ for the bulge component.

\new{ To identify BHs with galaxies for analysis, we combine two spatial criteria: to be assigned to a galaxy, a BH must be located within 10\% of the galaxy's DM host halo virial radius, and simultaneously within two effective radii of the galaxy. Galaxy effective radii are computed by taking the geometric mean of the half-mass radius of the projected stellar densities along each of the simulation's Cartesian axes.} If several BHs meet both criteria for a given galaxy, the most massive BH is retained \citep[see][for details]{volonteri16} \new{and all other (i.e. secondary or wandering) BHs are discarded from the dataset analysed here.}. Galaxy mergers were identified using the DM halo merger trees constructed from the halo catalogues for $z<6$, which are built from snapshots that are spaced on average every 130 Myr. Galaxy mergers are classified into major (stellar mass ratios $> 1 : 4$) and minor (stellar mass ratios $1 : 4$ to $1 : 10$) galaxy mergers \citep[see][for details]{martin18}. During a BH merger, the less massive BH is considered to merge into the more massive one, whose identifier is retained. 

We identified a sample of 6851 galaxies at $z=0.0556$, (the average redshift of the observed merger-free `bulgeless' galaxy sample of \citealt{ssl17} for ease of comparison) from the Horizon-AGN simulation, which had central SMBH and for which bulge masses were computed. From this sample of galaxies, we selected sub-samples using three different criteria:
\begin{enumerate}
    \item A galaxy merger-based criterion, which selects galaxies according to the total number of galaxy mergers (both major, mass ratios $> 1 : 4$, and minor, mass ratio $ 1 : 10 - 1 : 4 $) since redshift $z=2$. The threshold of $z=2$ is motivated by the need for hierarchical structure formation at earlier epochs, as per $\Lambda$CDM, and is observationally motivated by the studies of \cite{martig12, martin18} who showed that galaxies with low bulge-to-total ratios have had no major mergers since at least $z \sim 2$. Galaxy mergers were identified using the DM halo merger tree in Horizon-AGN. We combined the number of major and minor mergers into a single parameter, $N_{\rm{merger}}$, to quantify the total number of mergers a galaxy has undergone since $z=2$. We use this to select a sample of galaxies which have been evolving in isolation with $N_{\rm{merger}} =0$, which resulted in 1781 ($26\%$) galaxies. We compare these to galaxies which have had their history dominated by galaxy mergers, with $N_{\rm{merger}} \geq 3$ (2117 galaxies, $\sim31\%$). 
    \item A BH merger-based criterion, which applies cuts to $f_{\rm BH,merge}$, the cumulative mass fraction of a BH gained through BH mergers. For an equal mass BH-BH merger, $f_{\rm BH,merge} = 0.5$ while for a $1:4 $ mass ratio merger, $f_{\rm BH,merge} = 0.2$. $f_{\rm BH,merge}$ can exceed 0.5 if a BH undergoes repeated mergers. No lower mass ratio cut is applied to mergers tracked in this way. We use this to select a sample of SMBHs which have evolved with minimal BH mergers with $f_{\rm BH,merge} < 0.1$, which resulted in 2137 ($\sim31\%$) galaxies. Note that 1609 galaxies in the simulation had \fmerge$=0$, with no mass contribution from mergers, and $3508$ galaxies ($\sim 51\%$ of the Horizon-AGN population at $z=0.0556$) have \fmerge$<0.25$ and so have had their SMBH growth dominated by non-merger evolution. We note that while there is a link between galaxy mergers and SMBH mergers, one of the reasons that the galaxy-merger based sample and the SMBH-merger based sample are not identical is that numerical SMBH mergers can be significantly delayed in comparison to the merger of their host galaxies, even without considering further delays due to processes on scales that are unresolved in HORIZON-AGN \citep{Volonteri20}.
    \item An observationally motivated bulge-to-total ($B/T$) cut to produce a disk-dominated sample, based on the ratio between the galaxy bulge mass $M_{\rm bulge}$ and the total galaxy stellar mass $M_{*}$. We use this to select a sample of galaxies with assumed galaxy merger-free histories with $B/T < 0.1$ (as is done observationally, e.g. \citealt{simmons13, ssl17}) which resulted in 179 ($\sim3\%$) galaxies. This small sample size is a combination of the low numbers of truly disk-dominated galaxies produced in simulations (due to disk instability triggered bulge formation), and the poor resolution of the simulation outputs on which the mass decomposition is performed (leading to an overestimate of the bulge mass in the simulation).
\end{enumerate}

$81$ galaxies are found in each of the 3 samples (i.e. with $N_{\rm{merger}}=0$, $f_{\rm BH,merge} < 0.1$ and $B/T < 0.1$). $1346$ galaxies are found in both the galaxy merger-free and BH merger-free samples (i.e. $N_{\rm{merger}}=0$ and $f_{\rm BH,merge} < 0.1$).   Of the 179 galaxies selected to be disk-dominated, 112 ($63\%$) are classed as both disk-dominated and have not undergone a major or minor galaxy merger since $z=2$ (i.e. $N_{\rm{merger}}=0$ and $B/T < 0.1$). This suggests that the disk-dominated galaxies observed by \citet*{simmons13, ssl17}; \citet{smethurst19,smethurst21} are not a unique subset of galaxies, but are instead representative of the merger-free galaxy population.

\section{Results}\label{sec:results}

\begin{figure*}
    \begin{center}
	\includegraphics[width=\textwidth]{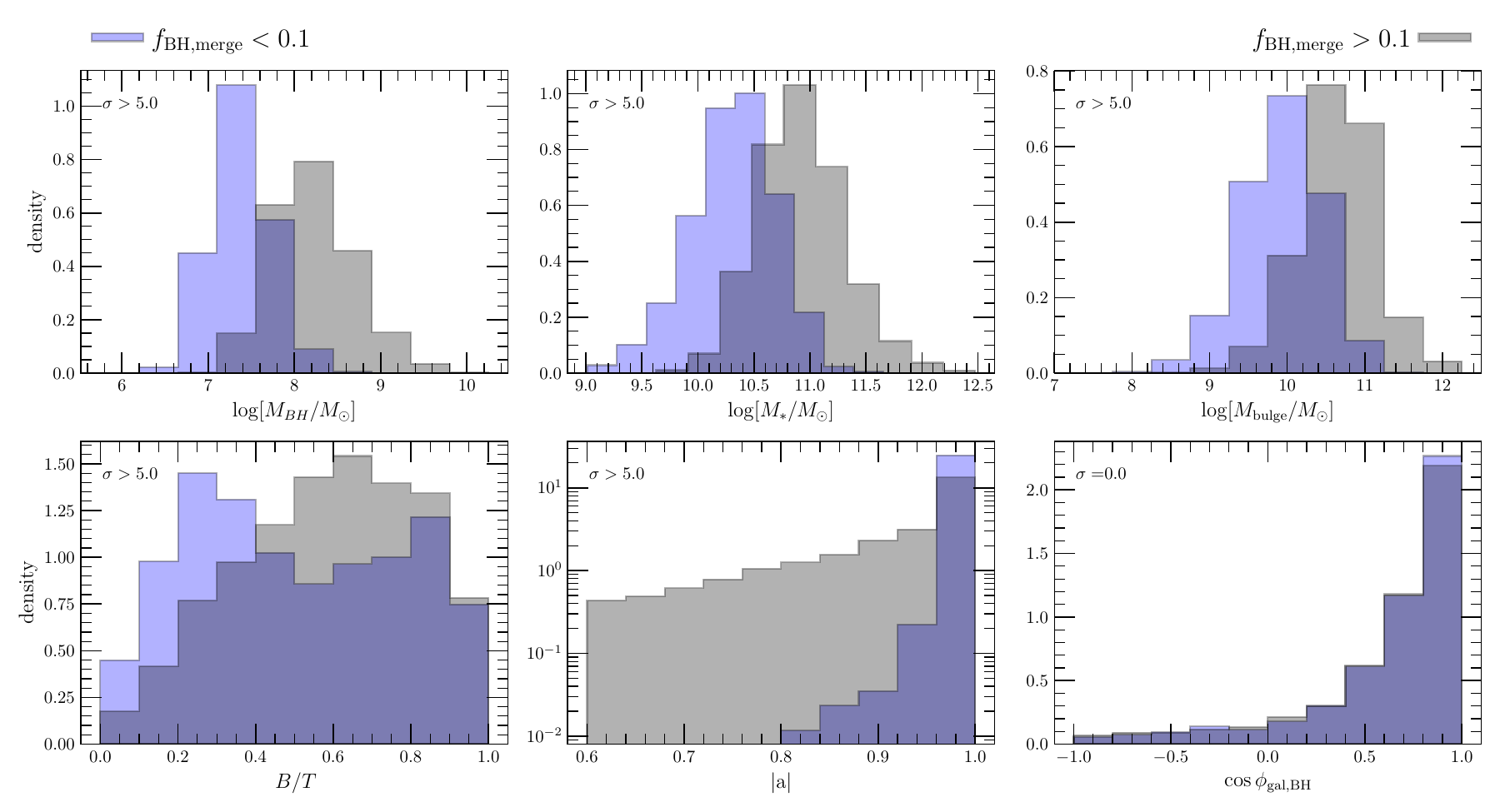}
	\vspace{-1em}
    \caption{The distribution of properties across the Horizon-AGN sample for galaxies selected to have low SMBH mass fractions grown by mergers, \fmerge$ <0.1$ (blue; and therefore accretion dominated SMBH growth), compared to the rest of the sample with \fmerge$>0.1$ (black; i.e. merger  and accretion grown SMBHs). Shown are the SMBH masses, $M_{BH}$, total stellar masses, $M_{*}$, bulge stellar mass, $M_{\rm{bulge}}$, the bulge-to-total stellar mass ratio $B/T$, the spin magnitude of the SMBH, $|a|$, and the angle between the spin of the SMBH and the galaxy, $\cos\phi_{\rm{gal, BH}}$. A value of $\cos\phi_{\rm{gal, BH}}=1$ means the SMBH and galaxy spins are aligned,  $\cos\phi_{\rm{gal, BH}}=0$ means they are misaligned by $90^{\circ}$, and $\cos\phi_{\rm{gal, BH}}=-1$ means they are misaligned by $180^{\circ}$. A value of $|a|=1$ represents a maximally spinning BH. Systems with SMBH mass fractions grown by mergers,  \fmerge$ < 0.1$, have statistically significantly lower SMBH masses, lower stellar masses and lower bulge masses. They have a bimodal distribution in bulge-to-total ratio, $B/T$, which is statistically significantly different to the rest of the population. They also have statistically significantly higher spin magnitudes, yet once again no statistically significant difference in spin alignment. We note that these results hold when incomplete, total stellar mass-matched samples of merger-free and merger-dominated systems are compared.}
    \label{fig:fmergesplit}
    \end{center}
\end{figure*}

\begin{figure*}
    \begin{center}
	\includegraphics[width=\textwidth]{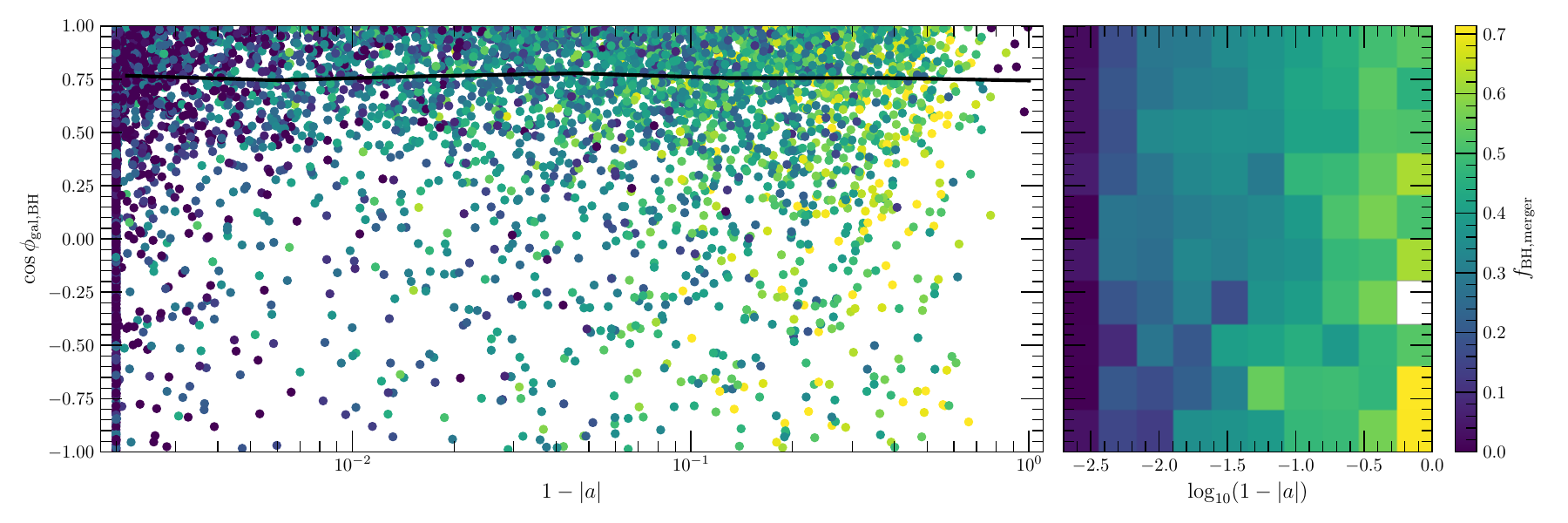}
	\vspace{-1em}
    \caption{The SMBH spin magnitude, $1-|a|$, plotted against the angle between the SMBH spin vector and galaxy spin vector, $\cos\phi_{\rm{gal,BH}}$, for our entire Horizon-AGN sample. In the left panel the data is plotted as a scatter plot with the points coloured by the fraction of the SMBH built by mergers, \fmerge~, with the black line showing the change in the average alignment with SMBH spin magnitude. In the right panel the data is binned and the median \fmerge~ value is shown for each bin. While there appears to be no obvious correlation between SMBH spin and alignment in the left panel, the right panel reveals that the highest \fmerge~ values are found at the lowest spin in the most misaligned systems (either perpendicular to the galaxy spin, or anti-aligned). Note that a low value of $1-|a|$ corresponds to a maximally spinning SMBH. We show $1-|a|$ here, as opposed to $|a|$, on a logarithmic scale to better appreciate the range of spins in maximally spinning systems.  A value of $\cos\phi_{\rm{gal, BH}}=1$ means the SMBH and galaxy spins are aligned,  $\cos\phi_{\rm{gal, BH}}=0$ means they are misaligned by $90^{\circ}$, and $\cos\phi_{\rm{gal, BH}}=-1$ means they are anti-aligned by $180^{\circ}$. We note that these results hold when incomplete, total stellar mass matched samples of merger-free and merger dominated systems are compared.}
    \label{fig:spinalign}
    \end{center}
\end{figure*}

\begin{figure*}
    \begin{center}
	\includegraphics[width=\textwidth]{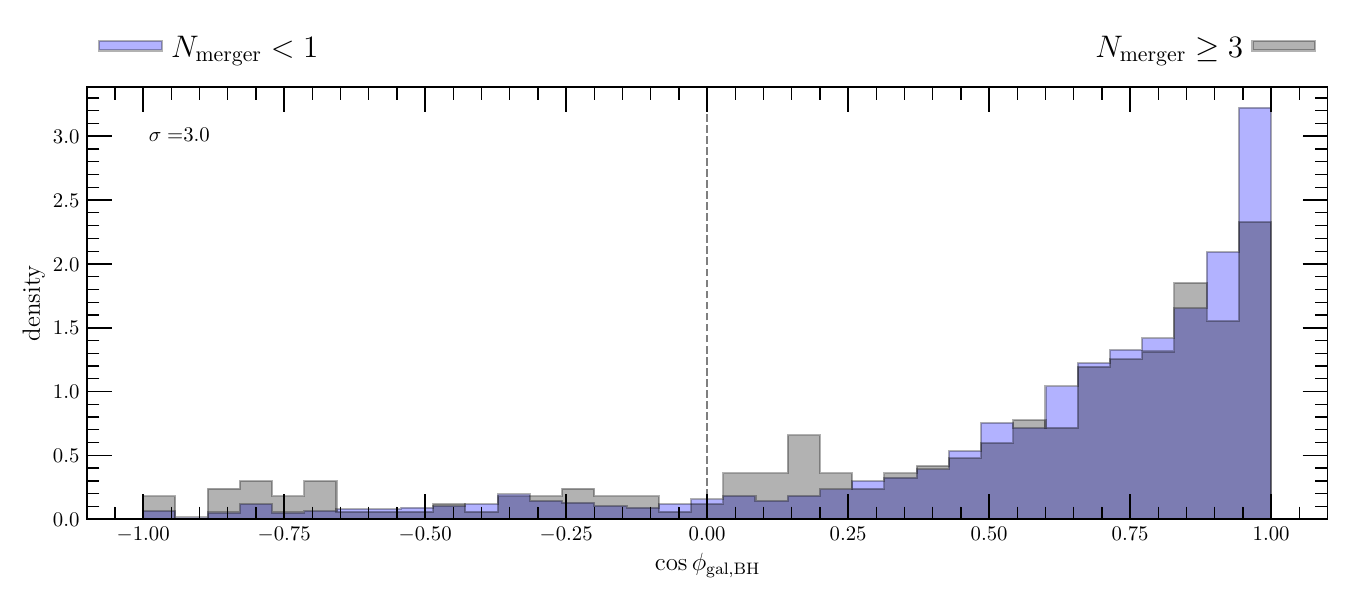}
	\vspace{-1em}
    \caption{The distribution of the angle between the galaxy and BH spin vector, $\cos\phi_{\rm{gal,BH}}$, for galaxies with SMBHs with less than $10\%$ of their mass due to mergers (\fmerge$>0.1$) and with more than $60\%$ of their mass due to mergers (\fmerge$>0.1$). These are the two extremes of the population, with distributions that are statistically significantly different ($3.4\sigma$). SMBHs dominated by non-merger growth are more likely to be aligned to their galaxy spins than SMBHs dominated by merger growth. A value of $\cos\phi_{\rm{gal, BH}}=1$ means the SMBH and galaxy spins are aligned,  $\cos\phi_{\rm{gal, BH}}=0$ means they are misaligned by $90^{\circ}$ (marked by the dashed line), and $\cos\phi_{\rm{gal, BH}}=-1$ means they are misaligned by $180^{\circ}$.} \label{fig:phidist}
    \end{center}
\end{figure*}

Figs. ~\ref{fig:BTsplit}, \ref{fig:Nmergesplit} \& \ref{fig:fmergesplit} show the properties of galaxies with merger-free and merger-dominated evolutionary histories identified using our 3 different criteria: bulge-to-total ratio (B/T; Fig.~\ref{fig:BTsplit}), number of minor and major galaxy mergers since $z=2$ ($N_{\rm{merger}}$; Fig.~\ref{fig:Nmergesplit}), and the fraction of the SMBH mass gained through BH mergers (\fmerge; Fig.~\ref{fig:fmergesplit}). We investigate the differences in the distributions of SMBH mass ($M_{\rm{BH}}$), total stellar mass ($M_{*}$), bulge stellar mass ($M_{\rm{bulge}}$), bulge-to-total ratio ($B/T$), fraction of the SMBH mass gained through BH mergers (\fmerge), SMBH spin magnitude ($|a_{\rm{BH}}|$), and the angle between the SMBH spin and galaxy spin ($\phi_{gal,BH}$; see equation~\ref{eq:angle}).

The masses of non-merger grown SMBHs are lower than for the rest of the population. This is apparent when non-merger grown SMBHs are selected using an observational cut with galaxy B/T ratio and with the cuts that simulations make possible of number of mergers and fraction of SMBH mass grown by mergers ($>5\sigma$ statistical significance in a Kolmogorov–Smirnov test in all cases). This is also apparent for total and bulge stellar masses, with non-merger grown galaxies having lower stellar masses across Figures~\ref{fig:BTsplit}-\ref{fig:fmergesplit}. 

Non-merger grown SMBHs selected on galaxy B/T and number of galaxy mergers are confirmed to have low \fmerge~(see Figs.~\ref{fig:BTsplit}~\&~\ref{fig:Nmergesplit}). However those galaxies selected with \fmerge$<0.1$ do not  uniformly have low bulge-to-total mass ratios. The distribution in Fig.~\ref{fig:fmergesplit} is bi-modal, with a peak at low and high B/T ratio, either side of the peak for the rest of the population. This suggests that galaxies which have grown their SMBHs without BH mergers ($\lesssim10\%$ by mass) can have significant bulges, perhaps grown by disk instabilities, although we caution that at $\sim 1 \rm \ kpc$ resolution, disk instabilities are likely to suppressed in HORIZON-AGN\footnote{The lack of resolution within the vertical structure of the disk acts like an extra source of temperature in the disk which most likely prevents it from secularly barring.}. \new{Using a semi-analytic model,} \cite{parry09} show in the Millennium simulation that bulge growth is dominated by disk instabilities\footnote{\new{Bulge growth due to instabilities is computed in the Millenium simulation using an inequality quantifying the dynamics of the disk, which when satisfied leads to either partial or full (depending on the model used) collapse of the mass in the disk into a spheroid. Horizon-AGN, being a hydrodynamical simulation, natively follows the formation of bulges through distributed star formation within the galaxy, but is limited by its spatial resolution of $\Delta x = 1 \rm kpc$}.} for galaxies with total stellar masses $<10^{11}\rm{M}_{\odot}$, with mergers only dominating for the most massive galaxies. This leads to secular growth of both bulges and SMBHs, resulting in the distribution of B/T for non-merger grown systems seen in Fig.~\ref{fig:fmergesplit}. 

The distribution of the SMBH spin magnitudes are $>5\sigma$ different for each of the three selections of non-merger grown systems. It is most apparent for those galaxies selected using \fmerge$<0.1$; this is unsurprising since this traces the true amount of mass in a SMBH grown by BH mergers (since $B/T$ and $N_{\rm{merge}}$ are both proxies for merger SMBH growth). The spins of non-merger grown SMBHs are maximal (confirming the hypothesis of \citealt{smethurst19}, which built on the work of \citealt{npk12}), in agreement with the results of \citet{dubois14,bustamante19}. We note that these results hold when incomplete, total stellar mass-matched samples of merger-free and merger-dominated systems are also compared. 

However, the distributions of alignment between the SMBH spin and galaxy spin, $\phi_{\rm{gal,BH}}$ are not statistically significantly different between the non-merger selected galaxies and the rest of the population in either of Figs.~\ref{fig:BTsplit},~\ref{fig:Nmergesplit}~\&~\ref{fig:fmergesplit}. This  appears to contradict the secular feeding hypothesis of \citet[][]{smethurst19} and the results from \cite{bustamante19}. We investigated this further in Fig.~\ref{fig:spinalign}. Under the secular feeding hypothesis, SMBHs grown by non-merger processes should be spun up due to alignment between galaxy and SMBH spin, so should have maximal spin ($1-|a|\sim0$) and be aligned ($\cos\phi_{\rm{gal,BH}}=1$). While the majority of such systems display this behaviour there is still a large tail to misaligned and anti-aligned systems (bottom left corner of Fig.~\ref{fig:spinalign}) due to the fact that individual accretion episodes can realign BH spin when alignment timescales are short, which increases the scatter in $\phi_{\rm{gal,BH}}$ \citep[see Appendix C of][]{dubois14}. \new{For a model with longer accretion timescales, we would expect a larger discrepancy between BHs in non-merger systems, who would have time to align with their host galaxy, and those in merger-dominated system, where BH spin alignment would lag after any changes in galactic spin due to galaxy merger. In such a scenario, merger-grown systems would be expected to be spun down and misaligned with their galaxies due to the variation in angular momentum of accreted gas introduced by galaxy mergers.} In our sample, merger-dominated systems show a similar scatter in $\phi_{\rm{gal,BH}}$ values as non-merger grown systems, suggesting that secular processes dominate the evolution of SMBHs in the epochs between galaxy mergers. While it is evident that BH merger-grown BHs are spun down by the mergers in Fig.~\ref{fig:spinalign}, there is no correlation between $\phi_{\rm{gal,BH}}$ and BH spin magnitude $|a|$ (solid line). \new{The caveat is that the spatial resolution of $\Delta x = 1 \rm \ kpc$ under-resolves the angular momentum structure in the centre of the galaxy, let alone on scales closer to the event horizon of the BH. While \citep{Dubois14HAGN} have shown that the spin evolution of a BH is reasonably well converged on resolutions of $10 - 80 \rm \ pc$, the significantly lower resolution of Horizon-AGN means that we probably over-predict the alignment between galactic and BH spin. We also note that the BH spin evolution model used here treats each accretion event as independent. If the timescales for a newly formed accretion disc to be consumed are significantly longer than the timestep of the simulation this would artificially increase the scatter in Fig.~\ref{fig:spinalign}.}

The right panel of Fig.~\ref{fig:spinalign} shows that there are differences in spin and spin alignment with \fmerge, but they are gradual enough across the parameter space that the single-threshold distributions of Figs.~\ref{fig:BTsplit}, \ref{fig:Nmergesplit}~\&~\ref{fig:fmergesplit} aren't able to capture them. However, its clear in Fig.~\ref{fig:spinalign} the SMBHs with the highest average mass fraction grown by BH mergers, \fmerge, are found in the lowest spin, most misaligned bins. This is only apparent when controlling for \fmerge~ and so in Fig. \ref{fig:phidist} we once again look at the distribution of $\cos\phi_{\rm{gal,BH}}$ but for the true extremes of the population; those non-merger driven systems with \fmerge$ < 0.1$ (2137 systems) and \fmerge$>0.6$ (318 systems). Their distributions of $\cos\phi_{\rm{gal,BH}}$ are statistically significantly different ($\sigma=3.4$ in a KS test), with BH merger grown systems more likely to be misaligned ($\cos\phi_{\rm{gal,BH}}\sim0$) or anti-aligned ($\cos\phi_{\rm{gal,BH}}\sim-1$), due to the spin flips caused by the misaligned orbital angular momentum of SMBH during mergers. It is only when probing the extremes of the population that this is apparent. The rest of the SMBH population evolves with a mix of merger and non-merger histories, with non-merger processes dominating ($51\%$ of SMBH  have $f_{\rm{BH,merger}} <0.25$, i.e. have not undergone a major merger with a mass ratio  of at least 1:3 since $z\sim2$) where high spins and alignment are expected. We note that while BH-BH mergers could not happen without the galaxy mergers that deliver multiple BHs to the same galaxy, the effect discussed here is entirely driven by the re-alignment of BH spins during a BH-BH merger, not by the rearranging of the galactic spin during galaxy mergers. As can be seen in the bottom right panel of Fig. \ref{fig:Nmergesplit}, there is no noticeable difference in the distribution of alignment angles between BH and galaxy for merger-rich and merger-poor galaxies. In addition, BH spins for BHs in both samples of galaxies are highly aligned with their host galaxy spin. This result will be at least partially influenced by the limited resolution of Horizon-AGN, and should be treated as an upper limit: at 1 kpc of resolution, the angular momentum distribution in the centres of galaxies is insufficiently well resolved to significantly decouple from the larger galactic angular momentum.

\section{Discussion}
\label{sec:discussion}

\begin{figure}
    \begin{center}
	\includegraphics[width=0.5\textwidth]{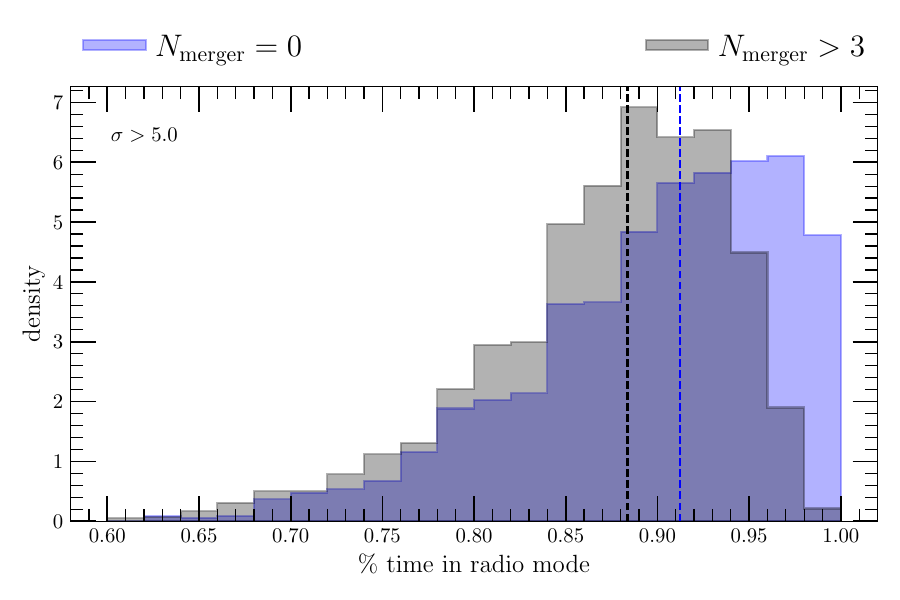}
	\vspace{-1em}
    \caption{The distribution of the time spent in a radio mode of AGN feedback (i.e. with $f_{\rm{Edd}}<1\%$) since $z=2$ for galaxies with merger-free evolutionary histories ($N_{\rm{merger}}=0$; blue) and those with merger dominated evolutionary histories ($N_{\rm{merger}}>3$; black). The median value for each distribution is shown by the dashed lines in the corresponding colours. The distributions are statistically significantly different ($>5\sigma$) with merger-free systems more likely to spend a larger proportion of their lifetimes in a radio mode of AGN feedback.} \label{fig:pctime}
    \end{center}
\end{figure}

\new{
\subsection{Timescales of spin alignment}
Our results have shown that SMBHs in merger-free galaxies have higher spins ($>5\sigma$) than the rest of the galaxy population and are more likely to be aligned with their galaxy spin ($3.4\sigma$). These results support the secular feeding hypothesis of \citet{npk12} and \cite{smethurst19} where material inflowing to the SMBH in the centre of a non-merger grown
system will come from within the galactic disk at a constant angular momentum vector, spinning up the SMBH to maximum, and subsequently align the spin of the SMBH according to that of the accretion disk formed from infalling gas through the Bardeen-Petterson effect \citep{bardeen75}.}

However, two populations of BHs remain difficult to explain in this hypothesis: (i) merger-grown SMBHs that are aligned with their galaxy but have low spin magnitudes, and (ii) merger-grown SMBHs with high spin magnitudes. Some of these effects will be statistical: for some BH-BH mergers, the spins of the two BHs and orbital angular momentum will roughly align, which means the spin of the post-merger BH will remain high. For others, the post-merger spin will happen to align closely with that of the galaxy. For aligned, low-spin BHs it is most likely a question of timescales: the timescale for a BH to re-align its spin vector with that of the host galaxy is much shorter than the characteristic timescale for it to increase its spin magnitude, as can be seen by that fact that on average all BHs in our sample, no matter their merger history, preferentially align with their host galaxy. A BH that has undergone a not-too-recent major merger, or series of more minor mergers, would only need a comparatively small amount of coherent accretion, possibly following a period of chaotic accretion during a galaxy merger that reduces its spin, to then realign itself with its host galaxy while retaining a low-spin.  The rate of this accretion would be too low to cause significant mass gain or spin-up the BH spin magnitude. Finally, systems might end up spinning highly if they are dominated by their orbital angular momentum: \cite{gammie04} argue that following a merger between two BHs, it is reasonable to assume that the final BH has an angular momentum that is equal to that of the binary. Specifically, `the merger of two BHs of comparable mass will immediately drive the spin parameter of the merged hole to 0.8' where the spin parameter is J/$\mathrm{M_{BH}}$.
\new{
\subsection{Spin alignment and the impact of AGN feedback}}

\new{Understanding the relation between the SMBH spin magnitude and alignment of the galaxy \& SMBH spin vectors in merger vs non-merger grown systems is crucial for our understanding of the impact of AGN feedback and therefore of galaxy evolution in its entirety.} 
AGN in Horizon-AGN are thought to be in one of two feedback modes. Those SMBHs with Eddington ratios, $f_{\rm Edd} < 1\%$ are thought to cause a `radio/kinetic/jet/maintenance mode' of AGN feedback, mostly affecting the galaxy halo. Whereas, those SMBHs with Eddington ratios, $f_{\rm Edd} > 1\%$ are thought to cause a  `quasar/thermal/radiative mode' of AGN feedback, heating the central regions of the galaxy during the peak luminosity of the AGN. In both modes, the radiative efficiency changes as a function of BH spin magnitude, with higher spinning BHs typically converting a larger fraction of their accreted mass into feedback energy. Given that we have shown that galaxy-merger-free grown SMBHs are more likely to have a higher spin, and therefore higher rotational energy, this could have important implications for the AGN feedback efficiency in such systems. 

Our results also suggest that such outflows from merger-free systems will be preferentially aligned with the galaxy spin and will therefore be ejected out of the plane of the galaxy if the AGN is in the directional `radio/kinetic/jet/maintenance' mode, rather than the isotropic `quasar/thermal/radiative' mode; such increased feedback efficiency may then only impact both the very centre of the galaxy and the galaxy halo, rather than progressing through the galactic disk. \citet{dave19} using the SIMBA simulation, found that the `radio/kinetic/jet/maintenance mode' of AGN feedback can drive population-wide galactic quenching from the `outside-in' as it heats the halo and cuts off the gas supply from the galactic halo to the disk. This quenches galaxies even in the absence of the more energetic `quasar/thermal/radiative mode' of AGN feedback which heats the surrounding galactic regions around the SMBH. Similar results on the importance of quenching in the absence of galaxy mergers were reported for IllustrisTNG \citep{Weinberger18,Xu22} and Horizon-AGN \citep{dubois2016,beckmann17,martin18}.

This difference in spin direction will only impact where the AGN feedback energy is delivered if the AGN spend a significant fraction of their time in `radio/kinetic/jet' mode. To understand whether this is the case, we investigated the SMBH Eddington ratios in those galaxies with merger-free ($N_{\rm{merger}}=0$) and merger-dominated ($N_{\rm{merger}}>3$) evolutionary histories. We calculated the fraction of time since $z=2$ that SMBHs spend in a radio mode of AGN feedback (i.e. with $f_{\rm{Edd}}< 10^{-2}$). The distributions for each sample are shown in Fig.~\ref{fig:pctime} along with the median values shown by the dashed lines. The two distributions are statistically significantly different (with $\sigma>5$ in a Kolmogorov–Smirnov (KS) test), with merger-free galaxies more likely to spend a higher proportion of their evolutionary histories in a radio mode of AGN feedback. Despite this, the averages for both samples remain high: SMBHs in merger-free galaxies spend, on average, $91\%$ of their lifetime in radio mode AGN feedback, while the sample in merger-dominated galaxies spends on average, $88\%$ of their lifetime in radio mode AGN feedback. While in radio mode, AGN drive powerful jets (here defined as $10^{-4} < f_{\rm edd} < 10^{-2}$) for a significant fraction of their time: \new{64 \% for SMBHs in merger-free galaxies and 59 \% for SMBHa in merger-dominated galaxies. This similarity does not strongly depend on the exact threshold value of $f_{\rm edd}$ chosen when transitioning from jet to quasar mode, with merger-free galaxies continuing to spend $\sim 5 $ percent more time in jet mode than merger-dominated galaxies for over an order of magnitude in difference in transition threshold.} This means that for a non-negligible fraction of their evolution, AGN feedback energy for merger-free galaxies is injected with a higher efficiency at a given accretion rate than for merger-grown SMBH due to the higher spin of SMBH in merger-free galaxies. This could potentially enhance the importance of radio-mode quenching \citep[as advocated by][]{dave19, Weinberger18} in merger-free galaxies, and compensate for the short-lived luminosity bursts of AGN post galaxy merger \citep{mcalpine11,volonteri16} which can lead to the effective quenching of the merger remnant \citet{dubois2016}. In addition, the similarity between the average time spent in a radio mode feedback for the merger-free and merger-dominated samples, again suggests that in the long epochs between galaxy mergers, merger-free accretion dominates the growth of SMBHs even in galaxy merger-rich systems and possibly makes an important contribution to their long-term co-evolution.

\new{
\subsection{Post-processed versus on-the-fly SMBH spin evolution}}

\begin{figure}
    \begin{center}
	\includegraphics[width=0.5\textwidth]{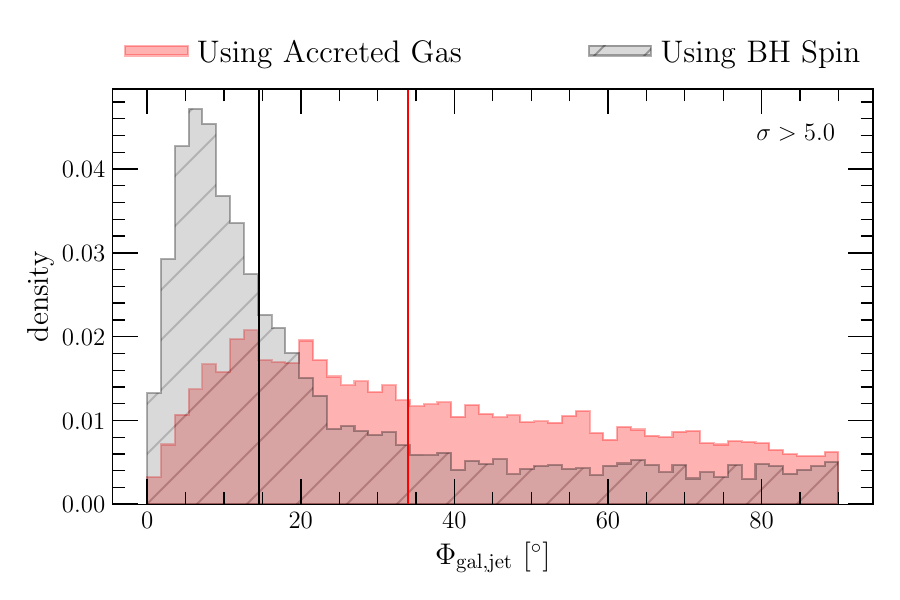}
	\vspace{-1em}
    \caption{\new{Distribution of $\Phi_{\rm gal,jet}$, the angle between the jet and
    the galactic angular momentum, for two jet models: a gas-based model where jets are injected along the angular momentum axis of gas accreted during the last accretion event (red solid) and a spin-based model where jets are injected along the spin axis of the SMBH (grey hatched). Median values for each distribution are shown by the solid lines of corresponding colours. As it disregards the inherent angular momentum of a spinning SMBH, the gas-based model (red solid) shows a higher degree of scatter and less long-term alignment between SMBH jets and galactic angular momentum than the spin-based model (grey hatched).} }\label{fig:pp_onthefly}
    \end{center}
\end{figure}

\new{One limitation of the work shown here is that the spin-evolution is post-processed from the BH mass evolution history, rather than self-consistently run on the fly. If the BH spin model had been run on the fly, AGN feedback in the simulation would have differed in the following two ways:}
\begin{enumerate}
\new{\item{Firstly, AGN jets would realign more slowly and coherently, as the BH spin vector is computed as an integrated quantity while the local gas angular momentum is an instantaneous quantity. This can be seen in Fig. \ref{fig:pp_onthefly}, which shows the angle between the galactic angular momentum and the jet for two cases: when using the instantaneous SMBH spin (spin-based model, post-processed), and using the angular momentum of the accreted gas for any given accretion and subsequent feedback event (gas-based model, which was used as the simulation was run). The gas-based model leads to less alignment between jets and the galactic angular momentum than the spin-based model. For disc galaxies, this would mean more jet energy is deposited directly into the circumgalactic medium in a spin-based than in a gas-based jet model. The difference in alignment between gas-based and spin-based models arises because in a spin-based model, the inherent angular momentum of the SMBH stabilises the jets against sudden realignment, while the direction of the central accreted angular momentum can be quite stochastic even for a well-ordered disc galaxy. Based on this insight, we would expect an on-the-fly spin-based model to drive jets into the circumgalactic more often than the current gas-based model, which would lead to a more indirect impact on galactic star formation, by cutting off large-scale gas flows, rather than direct quenching of star formation in galaxies.}}

\new{\item{Secondly, if SMBH spin were taken into account in BH’s radiative luminosity $\epsilon_r$, a given SMBH might produce anywhere from 0.3 to $\sim 4$ times the feedback energy at the same accretion rate in comparison to a fixed $\epsilon_r$. This would both change how much feedback energy the galaxy is receiving, and how fast the SMBH is growing (as the SMBH gains mass at a rate of $1- \epsilon_r$). For the model used here, feedback efficiency peaks for maximally spinning SMBH at a value of about 4 times the fixed $\epsilon_r = 0.1$ used to run the simulation, so we would expect secularly grown SMBHs to grow more slowly than those in merger-driven galaxies. This could reduce some of the effects of SMBHs in merger-free galaxies being overmassive in comparison to their host galaxy reported in the companion paper \citet{smethurst22}, or delay (but not prevent) the efficient spin-up of SMBHs in merger-free galaxies shown here in Fig. \ref{fig:fmergesplit}. The exact impact is hard to predict as SMBHs undergo self-regulation: strong feedback episodes reduce accretion onto the SMBHs in the short term, but long-term the balance between gas inflows and SMBH feedback might simply settle at a new equilibrium that produces a population of SMBHs with ultimately similar properties. How exactly a full on-the-fly model will different from post-processing will have to be tested using a follow-up simulation.}}
\end{enumerate}

\new{
\subsection{Follow-up observational work}

Given the predictions of our simulations presented here which suggest that the spin magnitude and orientation of galaxy-merger-free SMBHs are indeed different from the wider galaxy population and that this may lead to increased AGN feedback efficiency, high-resolution observational studies are therefore essential to test these predictions. Firstly, a study to test the alignment of AGN outflows with respect to their galactic disks in merger-free galaxies.} For example, using narrow band filters centered on \textsc{[oiii]} ionisation with the high spatial resolution provided by the Hubble Space Telescope to probe the alignment of AGN outflows. While previous studies have found no alignment between outflows and the galaxy minor axis (i.e. an alignment of SMBH spin and galaxy spin; see \citealt{kinney00, schmitt03, ruschel21}), such studies were of samples with mixed morphologies and therefore mixed evolutionary histories. By isolating merger-free systems, e.g. by observationally selecting bulgeless galaxies, and studying the orientation of their outflows we can observationally test the hypotheses discussed here.  

{\new Secondly, an observational study on the subsequent impact of the outflow on the galaxy star formation rate is required.} A high spatial and spectral resolution integral field unit, such as MUSE or VIRUS, will be able to spatially resolve the areas of disk-dominated merger-free galaxies impacted by AGN outflows and spectrally separate the emission ionised by star formation and the  outflow. Such a study would allow us to observationally test the assertions of \cite{dave19}  and the hypothesis discussed here: whether merger-free powered radio mode AGN feedback could indeed be the cause of galaxy population wide quenching.

\citet{smethurst21} found that their observational sample of 4 `bulgeless' (assumed merger-free) AGN with outflows, had outflow rates, energy injection rates and momentum fluxes which were comparable to a sample of low-$z$ Type 1 AGN from \citet[][within the scatter]{RW18}. \citeauthor{smethurst21} suggested that this result implied that it is possible that the majority of low-redshift AGN (both SMBH growth and outflows) are powered by non-merger processes. Given our results above, this suggests that the majority of this low-$z$ AGN feedback will occur via a radio mode. If future observational studies reveal that this feedback is capable of causing galaxy quenching, this could explain why a correlation is still observed between e.g. total stellar mass and SMBH mass for merger-free systems (see \citealt{simmons13, ssl17} and companion paper \citealt{smethurst22}); co-evolution regulated by radio mode AGN feedback is occurring due to secular processes. 

\section{Conclusions}\label{sec:conclusions}

We have investigated the SMBH spin magnitude and spin alignment in the context of the spin of their host galaxy for merger-free and merger-dominated galaxies using the Horizon-AGN simulation. Our conclusions are summarised as follows:
\begin{enumerate}
    \item Galaxies which have evolved in the absence of mergers host SMBHs with preferentially higher spin magnitudes than those with merger-dominated evolutionary histories. This is true for both SMBH mergers and galaxy mergers and supports the hypothesis of \citet{smethurst19}, building on the work of \citet{npk12}. 
    \item SMBHs with low mass fractions built by BH mergers (\fmerge $< 0.1$), as well as galaxies with few galaxy mergers, have a bi-modal distribution of bulge-to-total ratios (see Fig.~\ref{fig:fmergesplit} and Fig.\ref{fig:Nmergesplit}) suggesting that galaxies which have grown their SMBHs in the absence of mergers can still have significant stellar bulges, in agreement with \cite{parry09}. This leads to secular co-evolution of  both bulges and SMBHs.
    \item At first glance, SMBHs in galaxies with and without mergers have a very similar distribution of the angle between the spin vectors of the SMBH and their galaxy, $\phi_{\rm{BH,gal}}$ and show similar scatter in $\phi_{\rm{BH,gal}}$ (see Fig.~\ref{fig:spinalign}). It is only when probing the extremes of the population that it becomes apparent that the distribution of $\phi_{\rm{BH,gal}}$ are statistically significantly different ($3.4\sigma$) for SMBH dominated (\fmerge$>0.6$) or not (\fmerge$<0.1$) by BH mergers. This once again suggests that secular processes drive the evolution of SMBHs in the epochs between galaxy mergers, which for most objects dominates their overall evolution.
    \item Given that previous observational and theoretical works have also concluded that galaxy merger-free processes dominate SMBH-galaxy co-evolution, this suggests secular processes power the majority of the subsequent AGN feedback. We investigated the Eddington ratios of galaxies with merger-free and merger-dominated evolutionary histories, determining the time spent in a radio mode of AGN feedback (i.e. $f_{\rm{Edd}}<1\%$; see Fig.~\ref{fig:pctime}). We found that merger-free systems spend $91\%$ of their evolutionary history in a radio mode, similar to merger-dominated systems which spend $88\%$ of their time in radio mode, suggesting that the majority of AGN feedback occurs in this mode, in agreement with the findings from the SIMBA simulation \citep{dave19}. Given that we find that galaxy merger-dominated systems spend a similar average time in a radio mode, this once again suggests that in the epochs between galaxy mergers, merger-free accretion dominates the growth of their SMBHs.  If future observational studies reveal that this feedback is capable of causing galaxy quenching, this then suggests that the correlations between, e.g. total stellar mass and SMBH mass, seen for samples of merger-free galaxies \citep{simmons13, ssl17, smethurst22} is caused in part by secularly driven co-evolution regulated by radio mode AGN feedback. 
\end{enumerate}

High-resolution observational studies on the impact of AGN outflows and alignment with respect to their galactic disks in a pure sample of merger-free galaxies (e.g. disk-dominated galaxies) is therefore essential to both test these hypotheses of secularly powered AGN feedback which seems to dominate galaxy-SMBH co-evolution.    

\section*{Acknowledgements}

First authorship is shared between RSB and RJS. RSB assembled simulation data catalogues, interpreted results and wrote the manuscript. RJS conceived of the project, analysed data, interpreted results and wrote the manuscript. 

R. J. Smethurst gratefully acknowledges funding from Christ Church, Oxford and the Royal Astronomical Society. R. S. Beckmann gratefully acknowledges funding from Newnham College, Cambridge.  BDS acknowledges support from a UK Research and Innovation Future Leaders Fellowship [grant number MR/T044136/1]. This work is partially supported by grant Segal ANR-19-CE31-0017
of the French Agence Nationale de la Recherche.
 ILG acknowledges support from an STFC PhD studentship [grant number ST/T506205/1] and from the Faculty of Science and Technology at Lancaster University.

This work used the HPC resources of CINES (Jade supercomputer) under the allocation 2013047012 made by GENCI, and the horizon and Dirac clusters for post processing. This work is partially supported by the Spin(e) grants ANR-13-BS05-0002 of the French Agence Nationale de la Recherche and by the National Science Foundation under Grant No. NSF PHY11- 25915, and it is part of the Horizon-UK project, which used the DiRAC Complexity sys- tem, operated by the University of Leicester IT Services, which forms part of the STFC DiRAC HPC Facility (www.dirac.ac.uk). This equipment is funded by BIS National E-Infrastructure cap- ital grant ST/K000373/1 and STFC DiRAC Operations grant ST/K0003259/1. DiRAC is part of the National E-Infrastructure. This work has made use of the Horizon Cluster hosted by Institut d’Astrophysique de Paris. We thank Stéphane Rouberol for smoothly running this cluster for us. This research made use of Astropy,\footnote{http://www.astropy.org} a community-developed core Python package for Astronomy \citep{astropy13, astropy18}.

\section*{Data Availability}

All data used in this paper is available upon request to the first authors. 
 



\bibliographystyle{mnras}
\bibliography{refs.bib} 








\bsp	
\label{lastpage}
\end{document}